\documentclass[12pt]{article}
\usepackage{amsmath}
\usepackage{amsfonts}
\usepackage{amssymb}
\usepackage{eufrak}
\usepackage{graphicx}
\usepackage[mathscr]{eucal}
\begin{document}
\title{Critical behavior of the restricted primitive model}
\author{O.V.Patsahan, I.M.Mryglod\\
Institute for Condensed Matter Physics\\ 
1 Svientsitskii Str., 79011 Lviv, Ukraine \\
E-mail: oksana@icmp.lviv.ua
}
\maketitle

\begin{abstract}
We study  the critical behavior of the  systems dominated by Coulombic interaction.
 For this purpose we used  the method of collective variables  with a reference system.
Starting from the  Hamiltonian of the restricted primitive model (RPM), the simplest model of ionic fluids, we obtain the functional of the grand partition function given in terms of the two collective variables: the collective variables $\rho_{\mathbf{k}}$ and $c_{\mathbf{k}}$ describing fluctuations of the total number density and  charge density, respectively. As the result of  integration over the variables $c_{\mathbf{k}}$,   the microscopic based  effective  Hamiltonian of the RPM at the vicinity of its gas-liquid critical point was constructed.
The coefficients of the effective Hamiltonian, describing the density fluctuations nearby the gas-liquid critical point, are analyzed. It is shown that, in spite of the long-range character of the Coulombic potential, the effective interactions appeared at this level of the description  have the short-range character.   As a result, the effective Hamiltonian obtained for the RPM in the vicinity of the  critical point has the form of the Ginzburg-Landau-Wilson Hamiltonian of an Ising-like model  in a magnetic field.  This confirms the fact that the critical behavior of the RPM nearby the gas-liquid critical point belongs to the  universal class of a 3D Ising model.\\
\end{abstract}

\section{Introduction}
Experimental investigations of the critical properties of electrolyte solutions displayed three different types of behavior: Ising-like and mean-field behavior as well as a crossover between the two \cite{levelt1}-\cite{stell2}. In order to interpret the results, ionic systems were classified  in solvophobic and Coulombic. In solvophobic systems, Coulombic forces are not supposed to play a major role; the critical behavior is that common for fluids and fluid mixtures, i.e. Ising-like. By contrast, in Coulombic systems the phase separation is driven by electrostatic interactions. In recent years the critical behavior of the Coulombic systems has been a subject of active research. A theoretical model which demonstrates the phase separation driven exclusively by Coulombic forces is the restricted primitive model (RPM) \cite{fisher1,stell1}. In this model the ionic fluid is described as an equimolar-equisized mixture of opposite charged hard spheres immersed in a dielectric medium. Early studies established \cite{stillinger}-\cite{stellwularsen} that the model has a gas-liquid (GL) phase transition. A reasonable theoretical description of the GL critical point in the RPM was accomplished at a mean-field (MF) level using integral equation methods \cite{stell1,stell3} and Debye-H\"{u}ckel theory \cite{levinfisher}. However, a systematic theoretical investigation of the criticality of the RPM requires the microscopic based  effective Ginzburg-Landau-Wilson (GLW) Hamiltonian. Although some progress in this direction has been made  \cite{fisher3}-\cite{ciach1} the GLW functional of the continuous RPM explicitly related to the microscopic characteristics has not been derived yet.

For the last decade, the GL critical point in the RPM has been much studied
by computer simulation methods \cite{panagiotopoulos3}-\cite{kim}.  While the theory has not provided the clear picture of thermodynamic behavior of the model in the critical region,very recent simulation studies have found strong evidence for Ising universal class \cite{luijten,kim}.

In this paper we address an issue of the criticality in the RPM using the functional representation of the grand partition function in terms of the collective variables (CVs). The transformation from the individual to the collective coordinates (the Fourier transforms of the density fluctuations) is carried out via the corresponding Jacobian. Within the random phase approximation for the Jacobian one arrives at the Debye-H\"{u}ckel approximation for the free energy \cite{zubar}. First, the CVs were introduced for the description of the charged particle systems in the 1950s \cite{bohm}-\cite{jukh}. The further development of the CV method was related to the theory of phase transition of the second type proposed in \cite{yuk}-\cite{patyuk4}.

According to the basic idea of universality \cite{kadanoff}, different systems  which can be described by  similar effective GLW Hamiltonians (the Hamiltonians of the same symmetry)  demonstrate  the same critical behavior. Thus, the knowledge of the effective Hamiltonian of the system it is important in determing  its critical behavior. Our purpose here is to derive, from the first principles, the effective GLW Hamiltonian of the RPM in the vicinity of its GL critical point.

The layout of the paper is as follows. We give a functional representation of the grand partition function of the RPM in Section 2. In Section 3 we construct the effective Hamiltonian. On this basis, some conclusions concerning the critical behavior of the RPM nearby the GL critical point are made in Section 3.

\section{Functional representation of the grand \\ partition function of
the RPM}

The RPM consists of $N=N_{+}+N_{-}$ hard spheres of diameter $\sigma$
with $N_{+}$ carrying charges $+q$ and $N_{-}$ ($=N_{+}$) charges $-q$, in a medium
of dielectric constant $D$. The interaction potential of the RPM has the form
\begin{equation}
U_{\gamma\delta}(r) = \left\{\begin{array}{ll}
                     \infty &\mbox{if~~ $r<\sigma$}\\
                     \frac{q_{\gamma}q_{\delta}}{Dr}~~~~~&\mbox{if~~ $r\geq \sigma$}
                     \end{array}
              \right. , \quad  q_{i}=\pm q.
\label{a1.1}
\end{equation}
We start with the grand partition function for a two-component system
($\gamma,\delta=+,-$):
\begin{displaymath}
\Xi=\sum_{N_{+}\geq 0}\sum_{N_{-}\geq 0}\prod_{\gamma=+,-}
\frac{z_{\gamma}^{N_{\gamma}}}{N_{\gamma}!}
\int(d\Gamma)
\exp\left[-\frac{\beta}{2}\sum_{\gamma\delta}\sum_{ij}
U_{\gamma\delta}(r_{ij})\right],
\end{displaymath}
where $(d\Gamma)=\prod_{\gamma}d\Gamma_{N_{\gamma}}$,
$d\Gamma_{N_{\gamma}}=d\vec r_{1}^{\gamma}d\vec r_{2}^{\gamma}\ldots d\vec
r_{N_{\gamma}}^{\gamma}$ ($\gamma=+,-$)
is an element of the configurational space of the $\gamma$th species;
$z_{\gamma}$ is the fugacity of the $\gamma$th species:
$z_{\gamma}=\exp(\beta\mu_{\gamma}^{'})$,
$\mu_{\gamma}^{'}=\mu_{\gamma}+\beta^{-1}\ln[(2\pi m_{\gamma}\beta^{-1})^{3/2}/h^3]$, $\mu_{\gamma}$ is the chemical potential of the $\gamma$th species and
$\mu_{\gamma}^{'}$ is determined from the equation
\begin{displaymath}
\frac{\partial\ln\Xi}{\partial\beta\mu_{\gamma}} = \langle
 N_{\gamma}\rangle.
\end{displaymath}
For the RPM we have $\mu_{+}=\mu_{-}$.

Now we present the interaction potential $U_{\gamma\delta}(r)$ as a sum of two terms:
\[
U_{\gamma\delta}(r)=\psi_{\gamma\delta}(r)+\Phi_{\gamma\delta}(r),
\]
where $\psi_{\gamma\delta}(r)$  is a potential of a short-range repulsion and
$\Phi_{\gamma\delta}(r)$ is a long-range attractive part of the potential. We split the potential $U_{\gamma\delta}(r)$ into short- and long-range parts using the Weeks-Chandler-Andersen partition \cite{wcha}. As a result, we have
\begin{displaymath}
\psi_{\gamma\delta}(r) = \left\{\begin{array}{ll}
                     \infty &\mbox{if~~ $r\leq\sigma$}\\
                     0~~~~~&\mbox{if~~ $r> \sigma$}
                     \end{array}
              \right. ,
\end{displaymath}
\begin{displaymath}
\Phi_{\gamma\delta}(r) = \left\{\begin{array}{ll}
                     \frac{q_{\gamma}q_{\delta}}{D\sigma}&\mbox{if~~ $r\leq\sigma$}\\
                     \frac{q_{\gamma}q_{\delta}}{Dr}~~~~~&\mbox{if~~ $r> \sigma$}
                     \end{array}
              \right. .
\end{displaymath}

This simple form for $\Phi_{\gamma\delta}(r)$ inside the hard core changes the behavior of the
Fourier transform  for large $k$ from usual Coulombic $k^{-2}$ to $k^{-3}$ decay. As was shown
\cite{cha}, this choice of $\Phi_{\gamma\delta}(r)$ for $r<\sigma$ produces rapid convergence
of the series of the perturbation theory for the free energy.
The Fourier transform of $\Phi_{\gamma\delta}(r)=\frac{q^{2}}{Dr}=\Phi_{C}(r)$ has the form
\begin{equation}
\beta\rho\tilde \Phi_{C}(x)=24\beta^{*}\eta \frac{\sin x}{x^3},
\label{a2.1}
\end{equation}
where $\beta^{*}=\frac{\beta q^2}{D\sigma}$, $\beta=\frac{1}{k_{B}T}$, $\eta=\frac{\pi}{6}\rho\sigma^3$ is fraction density, $x=k\sigma$.

Within the framework of the approach considered the  interaction connected with a repulsion (potential $\psi_{\gamma\delta}(r)$) is  described in  the  space  of the Cartesian coordinates of the particles. We call the hard sphere system with the diameter $\sigma$ a  reference  system  (RS). The interaction connected with an attraction (potential $\Phi_{\gamma\delta}(r)$ ) is considered in the CV space. The transformation from the Cartesian coordinates to the CVs is performed by means of the transition Jacobian.

Using the method of CVs developed for a two-component continuous system \cite{oksana}-\cite{patkozmel}  we can rewrite the grand partition function  of the RPM in the following form:
\begin{equation}
\Xi=\Xi_{0}\int (d\rho)(dc)\exp\left(\beta\mu_{1}\rho_{0}-\frac{\beta}{V}\sum_{\mathbf{k}}
\tilde{\Phi}_{C}(k)c_{\mathbf{k}}c_{-\mathbf{k}}\right)J(\rho,c).
\label{a2.2}
\end{equation}
Here $\Xi_{0}$ is the grand partition function of the RS. $\rho_{\bf{k}}$ and $c_{\bf{k}}$ are
the CVs which describe  total density fluctuation modes and charge density fluctuation modes,
respectively:
$$
\rho_{\mathbf{k}}=\rho_{\mathbf{k}}^{c}-\mathrm{i}\rho_{\mathbf{k}}^{s}, \qquad
c_{\mathbf{k}}=c_{\mathbf{k}}^{c}-\mathrm{i}c_{\mathbf{k}}^{s},
$$
the indices $c$ and $s$ denote the real and  imaginary parts of CVs
$\rho_{\bf{k}}$ and $c_{\bf{k}}$. Each of $\rho_{\mathbf{k}}^{c}$ ($c_{\mathbf{k}}^{c}$)
and $\rho_{\mathbf{k}}^{s}$ ($c_{\mathbf{k}}^{s}$) takes all the real values from $-\infty$
to $+\infty$, and  $(d\rho)$ and $(dc)$ are volume elements of the CV phase space:
\[
(d\rho)=d\rho_{0}{\prod_{\bf{k}\neq 0}}^{'}d\rho_{\bf{k}}^{c}d\rho_{\bf{k}}^{s},
\qquad
(dc)=dc_{0}{\prod_{\bf{k}\neq 0}}^{'}dc_{\bf{k}}^{c}dc_{\bf{k}}^{s}.
\]

Unknown parameter $\mu_{1}$ ($\mu_{1}=(\mu_{1,+}+\mu_{1,-})/\sqrt{2}$)
is determined from the equation
\begin{equation}
\frac{\partial\ln \Xi_{1}}{\partial\beta\mu_{1}} = \frac{\langle N\rangle}{\sqrt{2}}.
\label{a2.3}
\end{equation}

$J(\rho,c)$ is the Jacobian of the transition to the CVs averaged over the RS. For the RPM,  $J(\rho,c)$ is of  the same form as that for the symmetrical binary fluid \cite{patkozmel}:
\begin{equation}
J(\rho,c)=\int (d\omega)\,(d\gamma)\exp\Big[\mathrm{i}2\pi\sum_{\bf{k}}
(\omega_{\bf k}\rho_{\bf k}+\gamma_{\bf k}c_{\bf k})+
\sum_{n\geq 1}\sum_{i_{n}\geq 0}D_{n}^{(i_{n})}(\omega,\gamma)\Big],\label{dA.7}
\end{equation}
\begin{eqnarray}
D_{n}^{(i_{n})}(\omega,\gamma)&=&\frac{(-\mathrm{i}2\pi)^{n}}{n!}\sum_{{\bf{k}}_{1}\ldots{\bf{k}}_{n}}
{\mathfrak{\bar{M}}}_{n}^{(i_{n})}(k_{1},\ldots,k_{n})\times\nonumber \\
&  &\gamma_{{\bf{k}}_{1}}\ldots\gamma_{{\bf{k}}_{i_{n}}}
\omega_{{\bf{k}}_{i_{n+1}}}\ldots\omega_{{\bf{k}}_{n}}\delta_{{\bf{k}}_{1}+\ldots +{\bf{k}}_{n}},
\label{dA.8}
\end{eqnarray}
\begin{equation}
{\mathfrak{\bar{M}}}_{n}^{(i_{n})}=\frac{{\mathfrak{M}}_{n}^{(i_{n})}}{\sqrt{2}^{n}},
\label{dA.9}
\end{equation}
and variable $\omega_{\mathbf{k}}$ ($\gamma_{\mathbf{k}}$) is conjugate  to CV $\rho_{\mathbf{k}}$
($c_{\mathbf{k}}$).
Index $i_{n}$  ($i_{n}=0,2,4,\ldots 2n$) is used to indicate the number of variables $\gamma_{\bf{k}}$
in the cumulant expansion (\ref{dA.8}). Cumulants ${\mathfrak{M}}_{n}^{(i_{n})}$
are expressed as  linear combinations
of the partial cumulants ${\mathfrak{M}}_{\gamma_{1}\ldots\gamma_{n}}$ and are presented
for $\gamma_{1},\ldots,\gamma_{n}=+,-$  and $n\leq 4$ in \cite{patyuk4}
(see Appendix~B in \cite{patyuk4}).

In (\ref{dA.7})-(\ref{dA.8})  the cumulants ${\mathfrak{M}}_{n}^{(i_{n})}$ with $i_{n}=0$ are connected
with the $n$th structure factors of the RS \cite{patyuk4}:
\begin{equation}
{\mathfrak{M}}_{n}^{(0)}=\langle N\rangle S_{n}.
\label{dA.9a}
\end{equation}
Cumulants with $i_{n}\neq 0$ can be expressed in terms of ${\mathfrak{M}}_{n}^{(0)}$ (see also formulae (4.8) in \cite{patyuk4}):
\begin{eqnarray}
{\mathfrak{M}}_{n}^{(2)}&=&{\mathfrak{M}}_{n-1}^{(0)}, \qquad
{\mathfrak{M}}_{n}^{(4)}=3{\mathfrak{M}}_{n-2}^{(0)}-2{\mathfrak{M}}_{n-3}^{(0)},
\nonumber \\
{\mathfrak{M}}_{n}^{(6)}&=&15{\mathfrak{M}}_{n-3}^{(0)}-30{\mathfrak{M}}_{n-4}^{(0)}+
16{\mathfrak{M}}_{n-5}^{(0)}.
\label{a2.4b}
\end{eqnarray}
In general, the dependence of
${\mathfrak{\bar{M}}}_{n}^{(i_{n})}({\bf{k}}_{1},\ldots,{\bf{k}}_{n})$ on
wave vectors ${\bf{k}}_{1},\ldots,{\bf{k}}_{n}$ is complicated. Since we are interested in the critical behavior, the small-$\mathbf{k}$ expansion of the cumulant  can be considered.
Hereafter we shall replace
${\mathfrak{\bar{M}}}_{n}^{(i_{n})}({\bf{k}}_{1}, \ldots,{\bf{k}}_{n})$ by their values in
long-wave length limit ${\mathfrak{\bar{M}}}_{n}^{(i_{n})}(0,\ldots,0)$.
Structure factors $S_{n}(0,\ldots,0)$ with $n>2$ can be obtained from $S_{2}(0)$ by means of
a chain of equations for correlation functions \cite{stell}.

Let us present  $J(\rho,c)$ as
\begin{eqnarray}
J(\rho,c)&=&\int (d\omega)\,(d\gamma)\exp\Big[\mathrm{i}2\pi\sum_{\bf{k}}
(\omega_{\bf k}\rho_{\bf k}+\gamma_{\bf k}c_{\bf k})+\nonumber \\
&&
\frac{(-i2\pi)^{2}}{2!}\sum_{\bf k}({\mathfrak{\bar{M}}}_{2}^{(0)}
\omega_{\bf k}\omega_{-\bf k}+{\mathfrak{\bar{M}}}_{2}^{(2)}\gamma_{\bf k}\gamma_{-\bf k})+
\nonumber\\
&&
\sum_{n\geq 3}\sum_{i_{n}\geq 0}D_{n}^{(i_{n})}(\omega,\gamma)\Big],
\label{dA.10}
\end{eqnarray}
where  $D_{n}^{(i_{n})}(\omega,\gamma)$ has the following form (up to $n=4$):
\begin{eqnarray*}
D_{n}^{(i_{n})}(\omega,\gamma)&=&\frac{(-\mathrm{i}2\pi)^{3}}{3!}
\sum_{{\mathbf{k}}_{1},{\mathbf{k}}_{2},{\mathbf{k}}_{3}}
({\mathfrak{\bar{M}}}_{3}^{(0)}\omega_{{\mathbf{k}}_{1}}\omega_{{\mathbf{k}}_{2}}
\omega_{{\mathbf{k}}_{3}}+3{\mathfrak{\bar{M}}}_{3}^{(2)}\omega_{{\mathbf{k}}_{1}}
\gamma_{{\mathbf{k}}_{2}}\gamma_{{\mathbf{k}}_{3}})
\delta_{{\mathbf{k}}_{1}+{\mathbf{k}}_{2}+{\mathbf{k}}_{3}}+
\nonumber \\
&&\frac{(-\mathrm{i}2\pi)^{4}}{4!}
\sum_{{\mathbf{k}}_{1},\ldots,{\mathbf{k}}_{4}}
({\mathfrak{\bar{M}}}_{4}^{(0)}\omega_{{\mathbf{k}}_{1}}\omega_{{\mathbf{k}}_{2}}
\omega_{{\mathbf{k}}_{3}}\omega_{{\mathbf{k}}_{4}}+6{\mathfrak{\bar{M}}}_{4}^{(2)}
\omega_{{\mathbf{k}}_{1}}\omega_{{\mathbf{k}}_{2}}
\gamma_{{\mathbf{k}}_{3}}\gamma_{{\mathbf{k}}_{4}}+
\nonumber \\
&&{\mathfrak{\bar{M}}}_{4}^{(4)}\gamma_{{\mathbf{k}}_{1}}\gamma_{{\mathbf{k}}_{2}}
\gamma_{{\mathbf{k}}_{3}}\gamma_{{\mathbf{k}}_{4}})\delta_{{\mathbf{k}}_{1}+\ldots
+{\mathbf{k}}_{4}}.
\end{eqnarray*}
In (\ref{dA.10})  the linear term is  eliminated by the shift $\rho_{\bf k}=\rho_{\bf k}^{'}+{\mathfrak{\bar{M}}}_{1}^{(0)}\delta_{{\mathbf{k}}}$ (the prime on $\rho_{{\mathbf{k}}}$ is omitted for clarity).

According to (\ref{dA.9})-(\ref{a2.4b}), ${\mathfrak{\bar{M}}}_{2}^{(0)}(k)=
\frac{\langle N\rangle}{2}S_{2}(k) $
and ${\mathfrak{\bar{M}}}_{2}^{(2)}=\frac{\langle N\rangle}{2}$, where $S_{2}(k) $ is a two-particle structure factor  of a one-component hard sphere system. In the Carnahan-Starling approximation we have
for $S_{2}(k=0)$:
$$
S_{2}(0)=\frac{(1-\eta)^{4}}{1+4\eta+4\eta^{2}-4\eta^{3}+\eta^{4}}.
$$
\section{The effective  Hamiltonian of the RPM in the vicinity of the GL critical point}

Because ${\mathfrak{\bar{M}}}_{2}^{(0)}(0)$ is the  positive and smooth function in the region under consideration and ${\mathfrak{\bar{M}}}_{2}^{(2)}$ is equal to constant, we can integrate in (\ref{dA.10}) over $\omega_{\bf k}$ and $\gamma_{\bf k}$ using the Gaussian density measures as  basic ones.
This integration can be performed by several ways.  For example,  using  the Euler equations  we can
determine $\omega_{\bf k}^{*}$ (and $\gamma_{\bf k}^{*}$) which provide the maximum for the functional in the exponent in (\ref{dA.10}):
\begin{equation}
\omega_{\bf k}^{*}=\frac{\rho_{-\bf k}}{-\mathrm{i}2\pi{\mathfrak{\bar{M}}}_{2}^{(0)}}+\ldots,
\qquad
\gamma_{\bf k}^{*}=\frac{c_{-\bf k}}{-\mathrm{i}2\pi{\mathfrak{\bar{M}}}_{2}^{(2)}}+\ldots.
\label{dA.13}
\end{equation}
As a result, we can present $\Xi$ in the form:
\begin{eqnarray}
\Xi&=&\Xi_{0}{\cal{C}}\int(d\rho)(dc)\exp\Big[a_{1}^{(0)}\rho_{0}-\frac{1}{2!}\sum_{\bf k}\left(a_{2}^{(0)}
\rho_{\bf k}\rho_{-\bf k}+a_{2}^{(2)}c_{\bf k}c_{-\bf k}\right)+
\nonumber \\
&&
\frac{1}{3!}\sum_{{\mathbf{k}}_{1},{\mathbf{k}}_{2},{\mathbf{k}}_{3}}\left(a_{3}^{(0)}\rho_{-\bf k_{1}}
\rho_{-\bf k_{2}}\rho_{-\bf k_{3}}+a_{3}^{(2)}\rho_{-\bf k_{1}}c_{-\bf k_{2}}c_{-\bf k_{3}}\right)
\delta_{{\mathbf{k}}_{1}+{\mathbf{k}}_{2}+{\mathbf{k}}_{3}}+
\nonumber \\
&&
\frac{1}{4!}\sum_{{\mathbf{k}}_{1},\ldots,{\mathbf{k}}_{4}}\left(a_{4}^{(0)}\rho_{-\bf k_{1}}
\rho_{-\bf k_{2}}\rho_{-\bf k_{3}}\rho_{-\bf k_{4}}+a_{4}^{(2)}\rho_{-\bf k_{1}}
\rho_{-\bf k_{2}}c_{-\bf k_{3}}c_{-\bf k_{4}}+\right.
\nonumber \\
&&
\left.a_{4}^{(4)}c_{-\bf k_{1}}c_{-\bf k_{2}}c_{-\bf k_{3}}c_{-\bf k_{4}}\right)
\delta_{{\mathbf{k}}_{1}+\ldots+{\mathbf{k}}_{4}}\Big]
\label{dA.14}
\end{eqnarray}
and super index $i_{n}$ indicates the number of variables at $a_{n}^{(i_{n})}$. Here
\begin{equation}
{\cal{C}}=\prod_{\bf k}\frac{1}{\pi{\mathfrak{\bar{M}}}_{2}^{(0)}}\prod_{\bf k} \frac{1}{\pi{\mathfrak{\bar{M}}}_{2}^{(2)}}\exp(\beta\mu_{1}{\mathfrak{\bar{M}}}_{1}^{(0)}),
\label{dA.15}
\end{equation}
and the coefficients $a_{n}^{(i_{n})}$, taking into account the first terms in (\ref{dA.13}), can be written as
\begin{equation}
a_{1}^{(0)}=\beta\mu_{1}, \qquad
a_{2}^{(0)}=-\frac{1}{{\mathfrak{\bar{M}}}_{2}^{(0)}}, \qquad
a_{2}^{(2)}=-\frac{1+2\frac{\beta}{V}\tilde{\Phi}_{C}(k){\mathfrak{\bar{M}}}_{2}^{(2)}}
{{\mathfrak{\bar{M}}}_{2}^{(2)}},
\label{dA.16}
\end{equation}
\begin{equation}
a_{3}^{(0)}=\frac{{\mathfrak{\bar{M}}}_{3}^{(0)}}{({\mathfrak{\bar{M}}}_{2}^{(0)})^{3}},
\qquad
a_{3}^{(2)}=\frac{3{\mathfrak{\bar{M}}}_{3}^{(2)}}{{\mathfrak{\bar{M}}}_{2}^{(0)}
({\mathfrak{\bar{M}}}_{2}^{(2)})^{2}},
\label{dA.17}
\end{equation}
\begin{equation}
a_{4}^{(0)}=\frac{{\mathfrak{\bar{M}}}_{4}^{(0)}}{({\mathfrak{\bar{M}}}_{2}^{(0)})^{4}},
\qquad
a_{4}^{(2)}=\frac{6{\mathfrak{\bar{M}}}_{4}^{(2)}}{({\mathfrak{\bar{M}}}_{2}^{(0)})^{2}
({\mathfrak{\bar{M}}}_{2}^{(2)})^{2}},
\qquad
a_{4}^{(4)}=\frac{{\mathfrak{\bar{M}}}_{4}^{(4)}}{({\mathfrak{\bar{M}}}_{2}^{(2)})^{4}}.
\label{dA.18}
\end{equation}

Another way of integration in (\ref{dA.10}) is outlined in Appendix.

First, we restrict our attention to the Gaussian approximation
which corresponds to neglecting the terms proportional to $\rho^{3}$,
$\rho c^{2}$,  etc. in the exponent of (\ref{dA.14}). As was shown \cite{oksana_rpm1}, this approximation
yields only the boundary of stability with respect to fluctuations of the charge density. The fact that the
RPM does not demonstrate the GL phase instability in this approximatuin is attributed to the
effects of charge-charge correlations not present at this level.  In order to obtain the GL spinodal curve we should take into consideration the terms of  the order higher than the second one \cite{ciach1, oksana_rpm1}.

Now we follow the programme proposed in \cite{patyuk4,oksana} for a two-component
fluid system. First, we separate the two types of variables: the essential variables
(which include the variable connected with the order parameter) and the
non-essential variables. Then, integrating over the non-essential variables with the
Gaussian density measure, we construct the
basic density measure (the GLW Hamiltonian) with respect to the essential variables.

For the RPM in the vicinity of the GL critical point the variable $\rho_{\bf k}$  (describing the  fluctuations of the total number density) turns out to be the essential variable \cite{patkozmel}. Thus, we can present
(\ref{dA.14}) as
\begin{eqnarray}
\Xi&=&\Xi_{0}{\cal{C}}\int(d\rho)\exp\Big[\bar{a}_{1}\rho_{0}+\frac{1}{2!\langle N\rangle}\sum_{\bf k}\bar{a}_{2}
\rho_{\bf k}\rho_{-\bf k} +\frac{1}{3!\langle N\rangle^{2}}
\times
\nonumber \\
&&
\sum_{{\mathbf{k}}_{1},{\mathbf{k}}_{2},{\mathbf{k}}_{3}}
\bar{a}_{3}\rho_{-\bf k_{1}}
\rho_{-\bf k_{2}}\rho_{-\bf k_{3}}\delta_{{\mathbf{k}}_{1}+{\mathbf{k}}_{2}+{\mathbf{k}}_{3}}+
\frac{1}{4!\langle N\rangle^{4}}\sum_{{\mathbf{k}}_{1},\ldots,{\mathbf{k}}_{4}}\bar{a}_{4}
\times
\nonumber \\
&&
\rho_{-\bf k_{1}}\rho_{-\bf k_{2}}\rho_{-\bf k_{3}}\rho_{-\bf k_{4}}
\delta_{{\mathbf{k}}_{1}+\ldots+{\mathbf{k}}_{4}}\Big],
\label{dA.19}
\end{eqnarray}
where
\begin{equation}
\bar{a}_{n}=a_{n}^{(0)}+\triangle a_{n}
\label{dA.19a}
\end{equation}
and $\triangle a_{n}$ are the corrections obtained as the result of integration over CVs $c_{\bf k}$:
\begin{eqnarray}
\triangle a_{1}&=&\frac{1}{\sqrt{2}\langle N\rangle}\sum_{{\mathbf{q}}}\tilde{G}(q),\nonumber \\
\triangle a_{2}&=& \frac{S_{3}}{S_{2}^{2}}\frac{1}{\langle N\rangle}
\sum_{{\mathbf{q}}}\tilde{G}(q)+\frac{1}{2\langle N\rangle}
\sum_{{\mathbf{q}}}\tilde{G}(q)\tilde{G}(\mid{\mathbf{q}}-{\mathbf{k}}\mid)+\ldots ,\nonumber \\
\triangle a_{3}&=& \frac{6S_{3}}{\sqrt{2}S_{2}^{2}}\frac{1}{\langle N\rangle}
\sum_{{\mathbf{q}}}\tilde{G}(q)\tilde{G}(\mid{\mathbf{q}}+{\mathbf{k_{1}}}\mid)+\ldots ,\nonumber \\
\triangle a_{4}&=&\frac{6S_{3}^{2}}{S_{2}^{4}}\frac{1}{\langle N\rangle}
\sum_{{\mathbf{q}}}\tilde{G}(q)\tilde{G}(\mid{\mathbf{k_{3}}}+{\mathbf{k_{4}}}-{\mathbf{q}}\mid)+\ldots,
\label{dA.20}
\end{eqnarray}
where
\begin{equation}
\tilde{G}(q)=\frac{1}{1+\beta\frac{\langle N\rangle}{V}\tilde{\Phi}_{C}(q)}
\label{dA.21}
\end{equation}
is the charge-charge structure factor of the RPM determined in the Gaussian approximation.
 It is worth noting that  $\tilde{G}(q)$ is of
the same form as the function $\tilde{G}_{\phi\phi}^{0}(k)$ introduced  in \cite{ciach1}.
In this study we restrict our attention to the GL critical point  where $\tilde{G}(q)$ remains a smooth function.

Now let us consider the coefficient $\bar{a}_{2}$. Expanding  this coefficient
at small $k$ one can readily see that  the linear term  vanishes. As a result, we obtain
\begin{equation}
\bar{a}_{2}=\bar{a}_{2,0}+\frac{1}{2}k^{2}\bar{a}_{2,2}+\ldots,
\label{dA.22}
\end{equation}
where
\begin{eqnarray}
\bar{a}_{2,0}&=&a_{2}^{(0)}+\frac{S_{2}}{S_{3}^{2}}\frac{1}{\langle N\rangle}\sum_{\mathbf{q}}\tilde{G}(q)
+\frac{1}{2\langle N\rangle}\sum_{\mathbf{q}}\tilde{G}(q)^{2}, \nonumber \\
\bar{a}_{2,2}&=&\frac{1}{2\langle N\rangle}\sum_{\mathbf{q}}\tilde{G}(q)\frac{\partial^{2}\tilde{G}(q)}{\partial q^{2}}.
\label{dA.23}
\end{eqnarray}
Inserting the expansion (\ref{dA.22}) into (\ref{dA.19}) one arrives at the effective Hamiltonian of the RPM in the neighbourhood of the GL critical point.
The coefficients of the Hamiltonian are given explicitly in (\ref{dA.16})-(\ref{dA.18}) and (\ref{dA.19a})-(\ref{dA.23}).  As is seen, the effective interaction appeared in  (\ref{dA.19}) has  a  short-range character.

\section{Conclusions}

In this paper we have studied the critical behavior of the RPM at the vicinity of  the GL
critical point.  For this purpose the method of CVs with a reference system was  used . First we obtain
the functional of the grand partition function given in terms of the two CVs: the CV $\rho_{\mathbf{k}}$ describing fluctuations of the total number density and the CV $c_{\mathbf{k}}$ describing fluctuations of the charge density.
As is known,  the  Gaussian approximation of the functional of the grand partition function of simple fluids and their mixtures produces the  qualitative picture of their phase behavior. In contrast, the full phase
diagram of the RPM cannot be obtained within the framework of this approximation. In order to describe the GL phase transition, the terms of the higher order should be taken into
account in the effective Hamiltonian \cite{ciach1,oksana_rpm1}. Actually,  the charge-charge correlations  causes the effective attraction  between the ions that, in turn, leads to
the GL phase transition in the RPM.

 After the integration over CV $c_{\mathbf{k}}$ with the Gaussian basic density measure we construct  the effective Hamiltonian in terms of CVs $\rho_{\mathbf{k}}$ connected with the order parameter (up to $\rho^{4}$). All the coefficients of the effective Hamiltonian consist of  two parts: the part depending  solely on the characteristics of the RS (through  the $n$-particle   structure factors of the RS in long-wave length limit) and the part of the mixed type. The latter has the form of an expansion in terms of the charge-charge structure factors and is obtained as the result of integration over CVs  $c_{\mathbf{k}}$.
Allowance for the charge-charge  correlations (through the integration over CVs $c_{\mathbf{k}}$) leads to the contribution $\triangle a_{2}$ to the  coefficient $a_{2}$ (at the second power of CVs $\rho_{\mathbf{k}}$) which describes the effective attraction of the short-range character.

 Finally,  the original Hamiltonian is mapped onto the GLW Hamiltonian of an Ising-like model  in a magnetic field. We conclude that the form of the effective Hamiltonian of the RPM confirms the fact that the critical behavior of the RPM nearby the GL critical point belongs to the  universal class of a 3D Ising model.The coefficients of the effective Hamiltonian are given in terms of the structure factors of the RS as well as in terms of the charge-charge structure factors. This demonstrates that  the GL phase transition  in the RPM is attributed to the effects of strong charge-charge correlations which take place at the low temperatures.
A more comprehensive analysis of  the coefficients will be done elsewhere.

\section*{Acknowledgement}
Part of this work was supported by the Fundamental Research Fund of the Ministry of Education and Sciences of Ukraine for support under Project No. 02.07/00303.

\section*{Appendix}
We can rewrite  (\ref{dA.10}) in the form:
\begin{eqnarray*}
J(\rho,c)&=&\int (d\omega)\,(d\gamma)\exp\Big[\mathrm{i}2\pi\sum_{\bf{k}}
(\omega_{\bf k}\rho_{\bf k}+\gamma_{\bf k}c_{\bf k})+\nonumber \\
&&
\frac{(-i2\pi)^{2}}{2!}\sum_{\bf k}({\mathfrak{\bar{M}}}_{2}^{(0)}
\omega_{\bf k}\omega_{-\bf k}+{\mathfrak{\bar{M}}}_{2}^{(2)}\gamma_{\bf k}\gamma_{-\bf k})\Big]\times \nonumber\\
&&\Big[1+{\cal{A}}+\frac{1}{2!}{\cal{A}}^{2}+\ldots\Big],
\end{eqnarray*}
where
\begin{equation}
{\cal{A}}=\sum_{n\geq 3}\sum_{i_{n}\geq 0}D_{n}^{(i_{n})}(\omega,\gamma)
\label{f.2}
\end{equation}
We substitute  into  (\ref{f.2}) the operators $\frac{1}{i2\pi}\frac{\partial}{\partial \rho_{\bf k}^{'}}$
and $\frac{1}{i2\pi}\frac{\partial}{\partial c_{\bf k}}$ for $\omega_{\bf k}$ and $\gamma_{\bf k}$,
respectively. As a result, we can write for $J(\rho,c)$
\begin{equation}
J(\rho,c)=\Big[1+{\cal{\hat{A}}}+\frac{1}{2!}{\cal{\hat{A}}}^{2}+\ldots\Big]\exp\left[ -\frac{1}{2}
\sum_{\bf{k}}\frac{\rho_{{\mathbf{k}}}\rho_{-{\mathbf{k}}}}{{\mathfrak{\bar{M}}}_{2}^{(0)}}
-\frac{1}{2}\sum_{\bf{k}}\frac{c_{{\mathbf{k}}}c_{-{\mathbf{k}}}}{{\mathfrak{\bar{M}}}_{2}^{(2)}}
\right],
\label{f.13}
\end{equation}
where
\[
{\cal{\hat{A}}}=\sum_{n\geq 3}\sum_{i_{n}\geq 0}\hat{D}_{n}^{(i_{n})}(\partial/\partial \rho_{\bf k},\partial/\partial c_{\bf k}).
\]
Substituting (\ref{f.13}) into (\ref{a2.2}), we obtain for $\Xi$:
\begin{eqnarray*}
\Xi&=&\Xi_{0}\exp(\beta\mu_{1}{\mathfrak{\bar{M}}}_{1}^{(0)})\int(d\rho)(dc)\exp
(\beta\mu_{1}\rho_{0})\Big[1+{\cal{\hat{A}}}+\frac{1}{2!}{\cal{\hat{A}}}^{2}+\ldots\Big]
\times
\nonumber \\
&&\exp\left[ -\frac{1}{2}
\sum_{\bf{k}}\frac{\rho_{{\mathbf{k}}}\rho_{-{\mathbf{k}}}}{{\mathfrak{\bar{M}}}_{2}^{(0)}}
-\frac{\beta}{2V}\sum_{\bf{k}}\frac{1 +2\tilde{\Phi}_{C}(k)
{\mathfrak{\bar{M}}}_{2}^{(2)}}
{{\mathfrak{\bar{M}}}_{2}^{(2)}}c_{{\mathbf{k}}}c_{-{\mathbf{k}}}\right].
\end{eqnarray*}
and after cumbersome algebra we can arrive at the expressions (\ref{dA.14})-(\ref{dA.18}).

\end{document}